\documentclass{PoS}
\title{
$\rho$ meson decay width from 2+1 flavor lattice QCD
}
\ShortTitle{
$\rho$ meson decay width from 2+1 flavor lattice QCD
}
\author{
\speaker{N.~Ishizuka}${}^{a,b}$\thanks{E-mail : ishizuka@ccs.tsukuba.ac.jp}
\,\,\, for PACS-CS Collaboration,
\\ \\
\llap{$^a$}
Graduate School of Pure and Applied Sciences,
University of Tsukuba, Tsukuba 305-8571, Japan
\\
\llap{$^b$}
Center for Computational Sciences,
University of Tsukuba, Tsukuba 305-8577, Japan
}
\FullConference{
XXIX International Symposium on Lattice Field Theory - Lattice 2011 \\
July 10-16, 2011\\
Squaw Valley, Lake Tahoe, California
}
\abstract{
We perform a lattice QCD study of the $\rho$ meson decay
from the $N_f=2+1$ full QCD configurations
generated with a renormalization group improved gauge action
and a non-perturbatively $O(a)$-improved Wilson fermion action.
The resonance parameters,
the effective $\rho\to\pi\pi$ coupling constant
and the resonance mass,
are estimated from the $P$-wave scattering phase shift
for the isospin $I=1$ two-pion system.
The finite size formulas are employed to calculate the phase shift
from the energy on the lattice.
Our calculations are carried out
at two quark masses,
    $m_\pi=410\,{\rm MeV}$ ($m_\pi/m_\rho=0.46$)
and $m_\pi=300\,{\rm MeV}$ ($m_\pi/m_\rho=0.35$),
on a $32^3\times 64$ ($La=2.9\,{\rm fm}$) lattice
at the lattice spacing $a=0.091\,{\rm fm}$.
We compare our results at these two quark masses
with those given in the previous works using
$N_f=2$ full QCD configurations and the experiment.
}
\begin{document}
%
%
\section{ Introduction }
Study of the $\rho$ meson decay
is a significant step for understanding
the dynamical aspects of hadron interactions with lattice QCD.
In the early stage of studies toward this direction
the transition amplitude $\langle \pi\pi|\rho\rangle$
extracted from the time behavior of
the correlation function
$\langle \pi(t)\pi(t) \rho(0) \rangle$
was used to estimate the decay width,
assuming that the hadron interaction
is small~\cite{rhd:TAMP:GMTW,rhd:TAMP:LD,rhd:TAMP:MM,rhd:TAMP:JMMU}.

A more realistic approach is a study
from the $P$-wave scattering phase shift
for the isospin $I=1$ two-pion system.
The finite size formulas
presented by L\"uscher in the center of mass frame~\cite{Lfm:L}
and extensions to non-zero total momentum frames~\cite{Lfm:RG,Lfm:ETMC}
are employed for an estimation of the phase shift
from an eigenvalue of the energy on the lattice.
The first study of this approach
was carried out by CP-PACS Collaboration
using $N_f=2$ full QCD configurations
($m_\pi=330\,{\rm MeV}$, $a=0.21\,{\rm fm}$,
$La=2.5\,{\rm fm}$)~\cite{rhd:SCPH:CP-PACS}.
After this work
ETMC Collaboration presented results
with $N_f=2$ configurations
at several quark masses
($m_\pi=290,330\,{\rm MeV}$ ($La=2.5\,{\rm fm}$),
 $m_\pi=420,480\,{\rm MeV}$ ($La=1.9\,{\rm fm}$),
$a=0.079\,{\rm fm}$)~\cite{rhd:SCPH:ETMC_1,rhd:SCPH:ETMC_2}.
Recently Lang {\it et al.} reported
results of high statistical calculations
on a single $N_f=2$ gauge ensemble
($m_\pi=266\,{\rm MeV}$, $a=0.124\,{\rm fm}$,
$La=1.98\,{\rm fm}$)~\cite{rhd:SCPH:LANG}.

In the present work
we extend these studies by employing
$N_f=2+1$ full QCD configurations
and working on a larger lattice volume.
Our calculations are carried out with
the gauge configurations previously generated
by PACS-CS Collaboration
with a renormalization group improved gauge action
and a non-perturbatively $O(a)$-improved Wilson fermion action
at $\beta=1.9$ on $32^3\times 64$ lattice
($a=0.091\,{\rm fm}$, $La=2.9\,{\rm fm}$)~\cite{conf:PACS-CS}.
We choose two subsets of the PACS-CS configurations.
One of them corresponds to the hopping parameters
$\kappa_{ud}=0.13754$ for the degenerate up and down quarks and
$\kappa_{s }=0.13640$ for the strange quark,
for which the pion mass takes
$m_\pi=410\,{\rm MeV}$ ($m_\pi/m_\rho=0.46$).
The other is at
$\kappa_{ud}=0.13770$ and $\kappa_{s }=0.13640$,
corresponding to $m_\pi=300\,{\rm MeV}$ ($m_\pi/m_\rho=0.35$).
All calculations are carried out on
the PACS-CS computer at Center for Computational Sciences,
University of Tsukuba.
Details of our calculations are presented in Ref.~\cite{rhd:PACS-CS}.

We note that
QCDSF Collaboration calculated
the scattering phase shifts for the ground state
in the center of mass frame at several quark masses.
($m_\pi=240-430\,{\rm MeV}$)~\cite{rhd:SCPH:QCDSF}.
They estimated the resonance parameters from these results,
assuming that the effective $\rho\to\pi\pi$ coupling constant
does not depend on the quark mass.
BMW Collaboration
presented their first preliminary results
with $N_f=2+1$ configurations
($m_\pi=200, 340\,{\rm MeV}$, $a=0.116\,{\rm fm}$)
at Lattice 2010~\cite{rhd:SCPH:BMW}.
%
%
\section{ Method }
In order to calculate
the $P$-wave scattering phase shift
for the isospin $I=1$ two-pion system at various energies
from a single full QCD ensemble,
we consider three momentum frames,
the center of mass frame (CMF),
the non-zero momentum frames
with total momentum ${\bf P}=(2\pi/L)(0,0,1)$ (MF1)
and ${\bf P}=(2\pi/L)(1,1,0)$ (MF2),
as carried out in the previous works by
ETMC~\cite{rhd:SCPH:ETMC_1,rhd:SCPH:ETMC_2}
and Lang {\it et al.}~\cite{rhd:SCPH:LANG}.
In the present work
we calculate the scattering phase shifts
for four irreducible representations :
${\bf T}_{1}^{-}$ in the CMF,
${\bf A}_{2}^{-}$ and ${\bf E}^{-}$ in the MF1, and
${\bf B}_{1}^{-}$ in the MF2.
The finite size formulas for these representations
are given in Refs.~\cite{Lfm:L,Lfm:RG,Lfm:ETMC}.

For the ${\bf T}^{-}_1$ and the ${\bf E}^{-}$ representation,
we only calculate the scattering phase for the ground state
in the present work.
The energy of the ground state
is much smaller than that of the excited state
on our gauge configurations.
Thus the energy of these states
can be extracted by a single exponential fit for
the time correlation functions of the $\rho$ meson.
We use the local $\rho$ meson operator for the sink
and a smeared operator
for the source as discussed later.

For the ${\bf A}_{2}^{-}$ and the ${\bf B}_{1}^{-}$ representation,
we also calculate the scattering phase shift
for the first excited state.
In order to extract the energies of
the lowest two state for these representations,
we use the variational method~\cite{method_diag:LW}
with a matrix of the time correlation function,
\begin{equation}
G(t) = \left(
\begin{array}{ll}
       \langle 0|\, (\pi\pi)^\dagger(t)\, \overline{(\pi\pi)}(t_s)\, |0\rangle
& \,\, \langle 0|\, (\pi\pi)^\dagger(t)\, \overline{\rho    }(t_s)\, |0\rangle  \\
       \langle 0|\,     \rho^\dagger(t)\, \overline{(\pi\pi)}(t_s)\, |0\rangle
& \,\, \langle 0|\,     \rho^\dagger(t)\, \overline{\rho    }(t_s)\, |0\rangle
\end{array}
\right)
\ ,
\label{eq:def_G}
\end{equation}
for each representation.
The energies are extracted
from two eigenvalues $\lambda_n (t)$ ($n=1,2$) of the matrix
$M(t) = G(t)\, G^{-1}(t_R)$
with some reference time $t_R$,
assuming that the lower two states dominate
the correlation function.

In (\ref{eq:def_G})
the operator $\rho(t)$ is given by
$\rho(t) = \sum_{j=1}^{3}\, p_j \cdot \rho_j({\bf p},t) / |{\bf p}|$,
where $\rho_j({\bf p},t)$ is
the local operator
for the neutral $\rho$ meson at the time slice $t$ with the momentum ${\bf p}$.
The momentum takes
${\bf p}=(2\pi/L)(0,0,1)$ for the ${\bf A}_{2}^{-}$ and
${\bf p}=(2\pi/L)(1,1,0)$ for the ${\bf B}_{1}^{-}$ representation.
Hereafter
we assume that the momentum ${\bf p}$
takes one of these two values depending on the representation.
$(\pi\pi)(t)$ is an operator
for the two pions with the momentum ${\bf 0}$ and ${\bf p}$,
which is defined by
\begin{equation}
  (\pi\pi)(t) = \frac{1}{\sqrt{2}}
    \left(    \pi^{+}({\bf 0},t_1) \, \pi^{-}({\bf p},t)
            - \pi^{-}({\bf 0},t_1) \, \pi^{+}({\bf p},t)
    \right) \times {\rm e}^{ m_\pi \cdot ( t_1 - t ) }
\ ,
\label{eq:pp_op_sink}
\end{equation}
where
$\pi^{\pm}({\bf p},t)$ is the local pion operator
with the momentum ${\bf p}$ at the time slice $t$.
The time slice of the pion with the zero momentum
is fixed at $t_1 \gg t$,
and the time slice of the other pion $t$ runs over the whole time extent.
An exponential time factor
in (\ref{eq:pp_op_sink}) is introduced
so that the operator
has the same time behavior as that of the usual Heisenberg operator,
{\it i.e.},
$
    \langle 0 | \, (\pi\pi)^\dagger (t)
  = \langle 0 | \, (\pi\pi)^\dagger (0) \, {\rm e}^{ - H t }
$,
with the Hamiltonian $H$.

Two operators
$\overline{(\pi\pi)}(t_s)$ and
$\overline{\rho}(t_s)$
are used for the sources in (\ref{eq:def_G}),
which are given by
\begin{eqnarray}
&&
\overline{(\pi\pi)}(t_s)
= \frac{1}{\sqrt{2}}
   \Bigl(   \pi^{+}({\bf 0},t_s) \pi^{-}({\bf p},t_s)
          - \pi^{-}({\bf 0},t_s) \pi^{+}({\bf p},t_s)   \Bigl)
\ ,
\label{eq:pp_op_source}
\\
&&
\overline{\rho}(t_s) =
  \frac{1}{N_\Gamma}
  \sum_{{\bf z}\in \Gamma}
     \frac{1}{\sqrt{2}}
      \Bigl(   \overline{U}({\bf z},t_s) \gamma_p U({\bf z},t_s)
             - \overline{D}({\bf z},t_s) \gamma_p D({\bf z},t_s)   \Bigl)
   {\rm e}^{ i {\bf p}\cdot{\bf z} }
\ ,
\label{eq:rho_op_source}
\end{eqnarray}
where $\gamma_p=\sum_{j=1}^{3} p_j \cdot \gamma_j /|{\bf p}|$.
The operator
$Q({\bf z},t_s)$ ($Q=U,D$)
is a smeared operator
for the up or the down quark
given by
$
  Q({\bf z},t_s) = \sum_{\bf_x} q({\bf x},t_s )
                   \cdot \Psi(|{\bf x} - {\bf z}|)
$,
where $q({\bf x},t_s)$ ($q=u,d)$ is
the up or the down quark
at the position ${\bf x}$ and the time $t_s$.
We adopt the same smearing function
$\Psi(|{\bf x}|)$ as in Ref.~\cite{conf:PACS-CS}.
This operator
is used after fixing gauge configurations
to the Coulomb gauge.
In (\ref{eq:rho_op_source})
a summation over ${\bf z}$ is taken
to reduce a statistical error
and
$\Gamma = \{\, {\bf z}\, |\, {\bf z}=(L/2)\cdot(n_1,n_2,n_3)
  \, , \, n_j = \mbox{$0$ or $1$}\, , \, N_\Gamma = 8 \, \}
$
is chosen in the present work.
The smeared operator
(\ref{eq:rho_op_source})
is also used to extract the energy of the ground state for
the ${\bf T}_1^-$ and
the ${\bf E}^{-}$ representation,
setting the momentum ${\bf p}={\bf 0}$ and ${\bf p}=(2\pi/L)(0,0,1)$,
respectively.

The periodic boundary conditions are imposed
for both spatial and temporal directions
in configuration generations.
We impose the Dirichlet boundary condition
for the temporal direction
in calculations of the quark propagators,
to avoid the unwanted thermal contributions
produced by propagating two pions
in opposite directions in a time.
For both quark masses,
we set the source operators
at $t_s=12$ to avoid effects from the temporal boundary,
and the zero momentum pion in the sink operator
$(\pi\pi)(t)$ in (\ref{eq:pp_op_sink}) at $t_1=42$.
The total number of configurations at $m_\pi=410\,{\rm MeV}$
is $440$ and that $m_\pi=300\,{\rm MeV}$ is $400$.
We calculate the quark contractions of $G(t)$ in (\ref{eq:def_G})
by the source method and the stochastic noise method
as in the previous work by CP-PACS~\cite{rhd:SCPH:CP-PACS}.
Details of the method of the calculations
are explained in Ref.~\cite{rhd:PACS-CS}.
%
%
\section{ Results }
%
\begin{figure}[t]
\includegraphics[width=74mm]{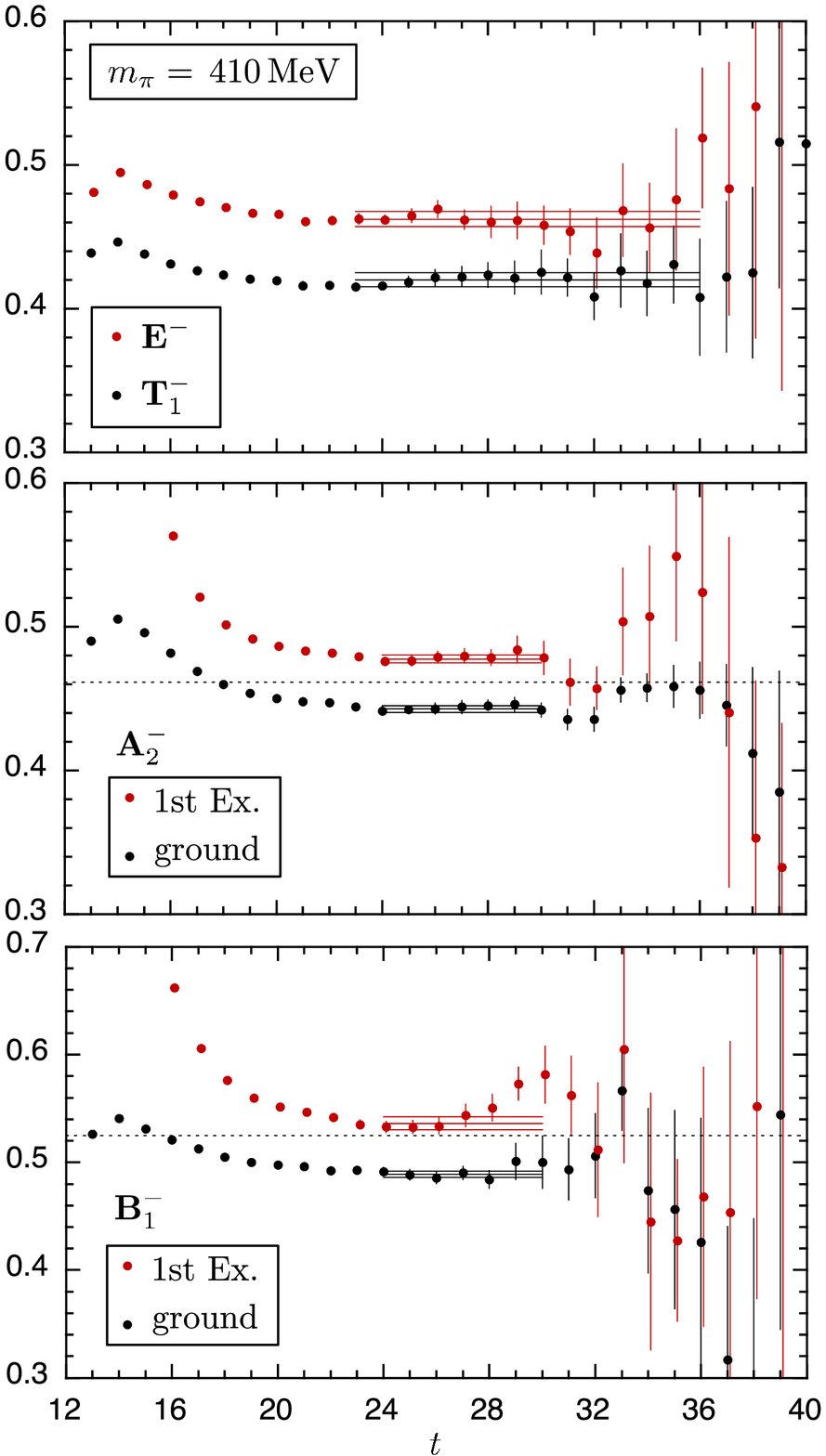}
\includegraphics[width=74mm]{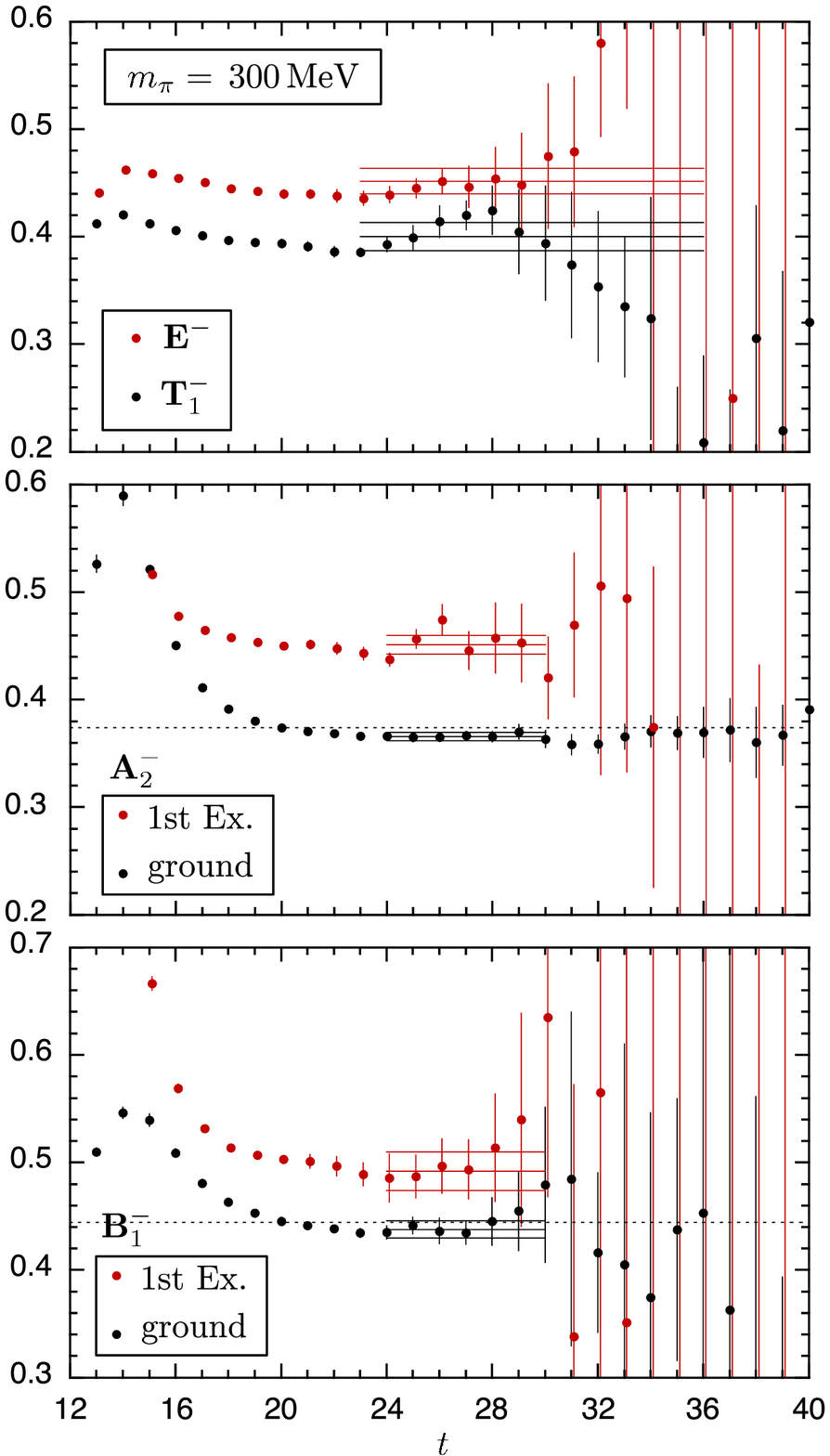}
\caption{
Effective masses
for six states considered in the present work
at
$m_\pi=410\,{\rm MeV}$ (left panel) and
$m_\pi=300\,{\rm MeV}$ (right panel).
}
\label{fig:LM}
\end{figure}
%
The effective masses of the time correlation functions
for six states considered in the present work
are plotted in Fig.~\ref{fig:LM},
where we choose $t_R=23$ as the reference time of the variational method.
We can find plateaus in the time region $t \ge 23$.
We extract the energy by a single exponential fit
for the time correlation functions.
In Fig.~\ref{fig:LM}
the results of the fitting
with one standard deviation error band
are also expressed by solid lines.
The dotted line for the ${\bf A}_2^-$
and ${\bf B}_1^-$ representation in the figure
indicates the energy of the two free pions
for each representation.

\begin{figure}[t]
\includegraphics[width=76mm]{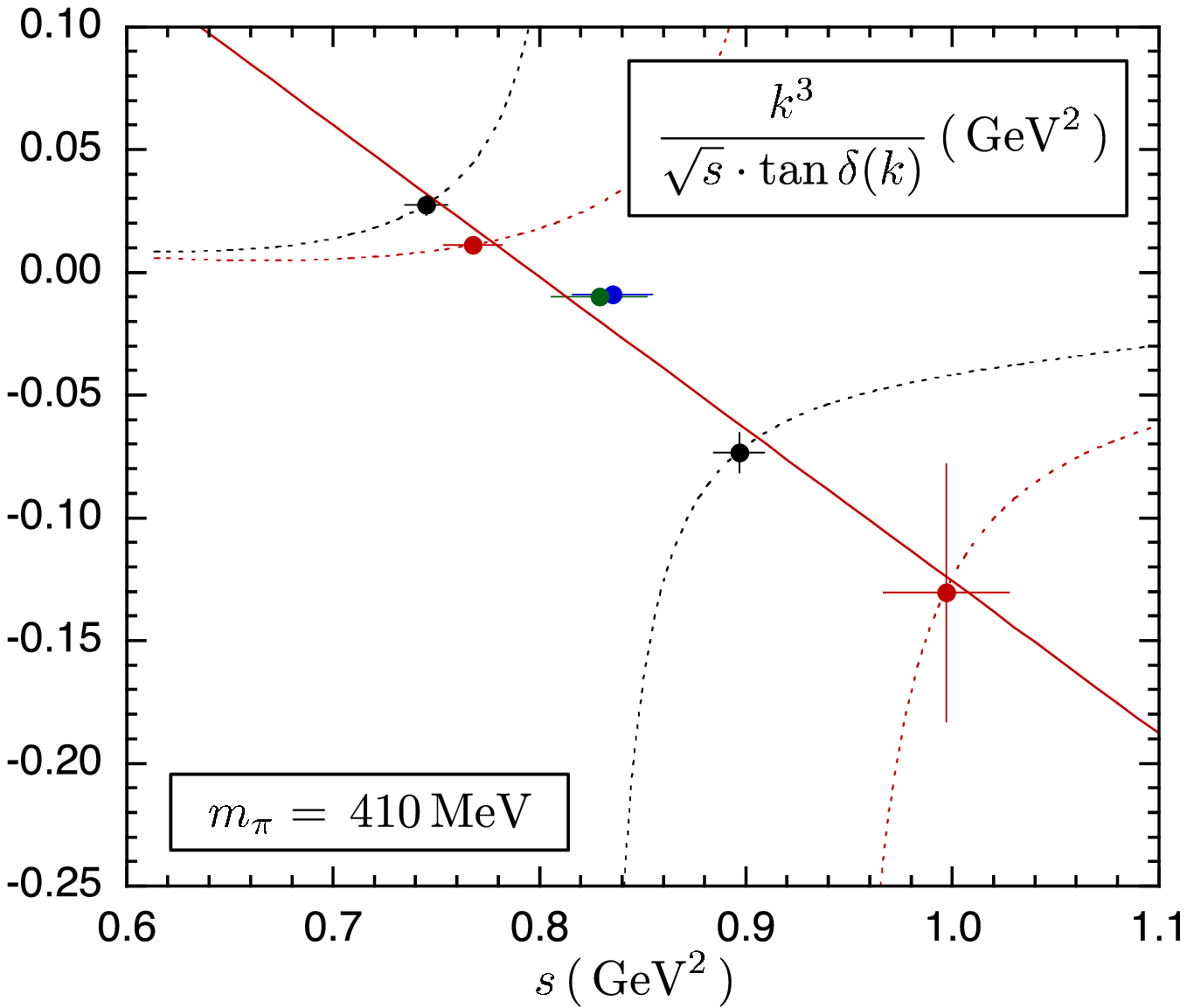}
\includegraphics[width=74mm]{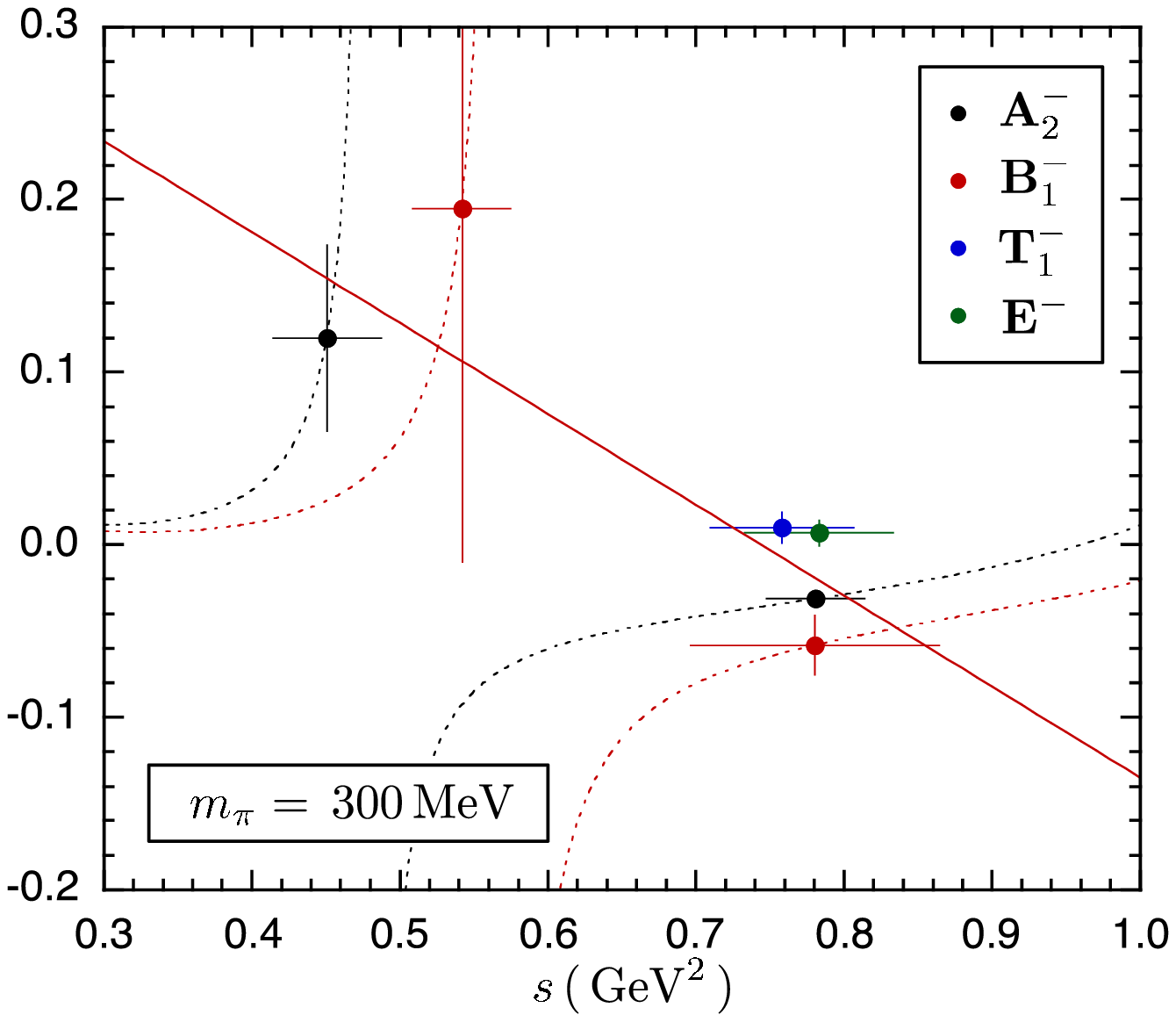}
\caption{
$(k^3/\tan\delta(k))/\sqrt{s}$
as a function of square of the invariant mass ${s}$
at  $m_\pi=410\,{\rm MeV}$ (left panel)
and $m_\pi=300\,{\rm MeV}$ (right panel).
}
\label{fig:k2_AMP_SS}
\end{figure}

Converting the energies for each representation
to the invariant masses $\sqrt{s}$
and substituting them into the finite size formulas,
we obtain the scattering phase shifts
plotted in Fig.~\ref{fig:k2_AMP_SS}.
In the figure
we show $( k^3 / \tan\delta(k) ) / \sqrt{s}$
as a function of square of the invariant mass $s$,
where $k=\sqrt{ s/4 - m_\pi^2 }$
is the scattering momentum.
The finite size formulas for the ${\bf A}_2^-$
and the ${\bf B}_1^-$ representation
are plotted by dotted lines.

In order to extract the resonance parameters from
our results of the scattering phase shift,
we parametrize the phase shift
with the effective $\rho\to\pi\pi$ coupling constant $g_{\rho\pi\pi}$
and the the resonance mass $m_\rho$ by
\begin{equation}
  \frac{k^3}{\tan\delta(k)} / \sqrt{s}
  = \frac{6\pi}{g_{\rho\pi\pi}^2}
    \,( m_\rho^2 - s )
\ .
\label{eq:k2_AMP_ERF}
\end{equation}
This parametrization has been widely used
in the previous works of the $\rho$ meson decay.
By chi-square fitting of the scattering phase shifts
with the fit function (\ref{eq:k2_AMP_ERF}),
we obtain,
\begin{eqnarray}
&&
 g_{\rho\pi\pi} = 5.52   \pm 0.40  \quad , \quad
   m_\rho       = 892.8  \pm 5.5  \pm 13  \,{\rm MeV}
\qquad \mbox{ at $m_\pi=410\,{\rm MeV}$ }
\ ,
\label{eq:result_410}
\\
&&
 g_{\rho\pi\pi} = 5.98  \pm 0.56         \quad , \quad
   m_\rho       = 863   \pm 23 \pm 12    \,{\rm MeV}
\quad
\qquad \mbox{ at $m_\pi=300\,{\rm MeV}$ }
\ ,
\label{eq:result_300}
\end{eqnarray}
where the second error of $m_\rho$
is the systematic uncertainty
for the determination of the lattice spacing.
In Fig.~\ref{fig:k2_AMP_SS}
we draw the fitting curves by solid red lines.
%
%

From (\ref{eq:result_410}) and (\ref{eq:result_300})
we find that the $g_{\rho\pi\pi}$ at the two quark masses
are consistent within the statistical error and also
with the experiment $g_{\rho\pi\pi}=5.874\pm 0.014$
given from the experimental results of the decay width
$\Gamma_\rho = 146.2 \pm0.7\,{\rm MeV}$~\cite{PDG:2010}.
This suggests a weak quark mass dependence
of the coupling constant.
But our calculations are carried out only at the two quark masses,
so calculations at more quark masses are necessary
to obtain a definite conclusion for the quark mass dependence.
We leave this issue to studies in the future.
%
\begin{figure}[t]
\includegraphics[width=74mm]{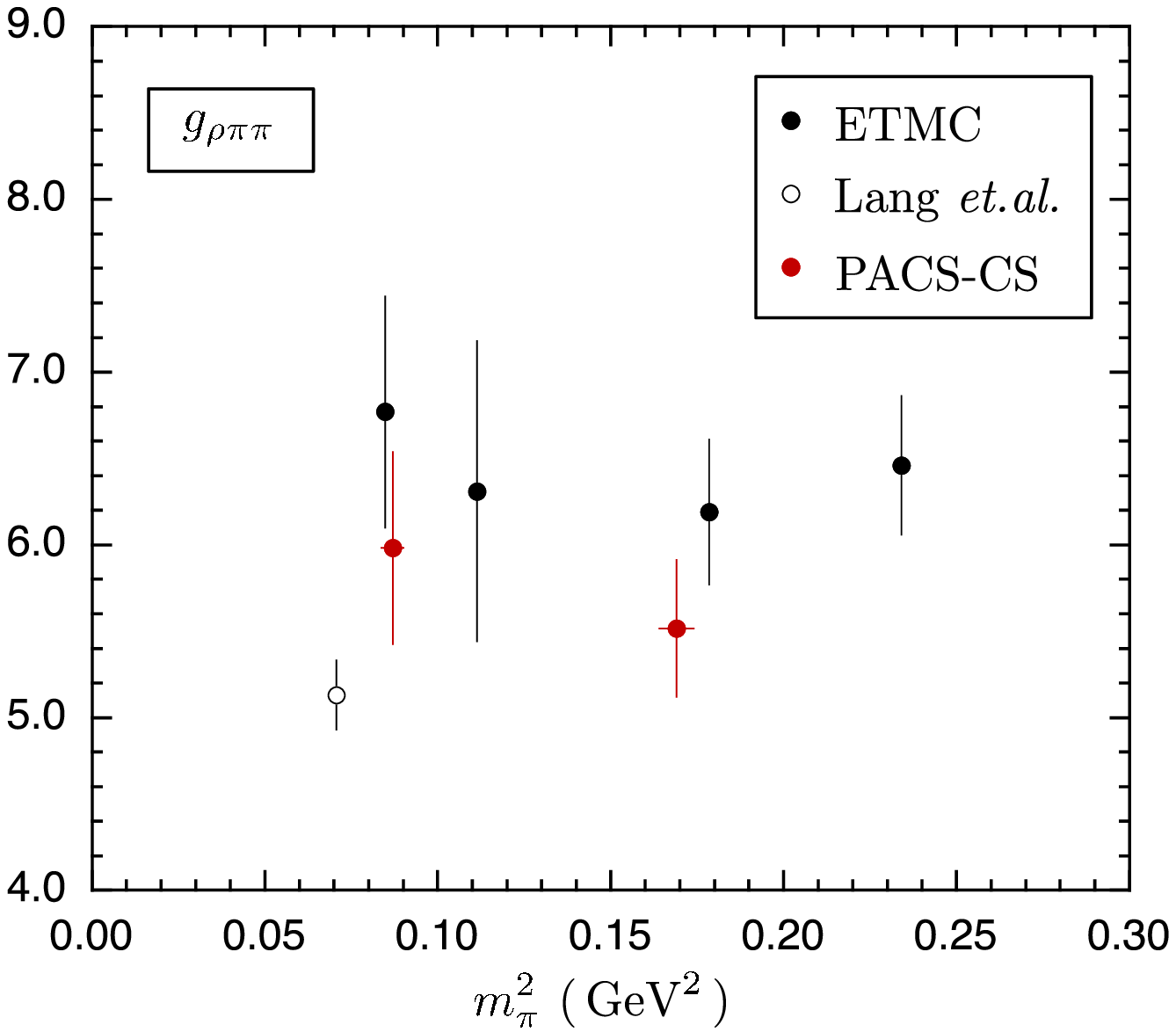}
\includegraphics[width=76mm]{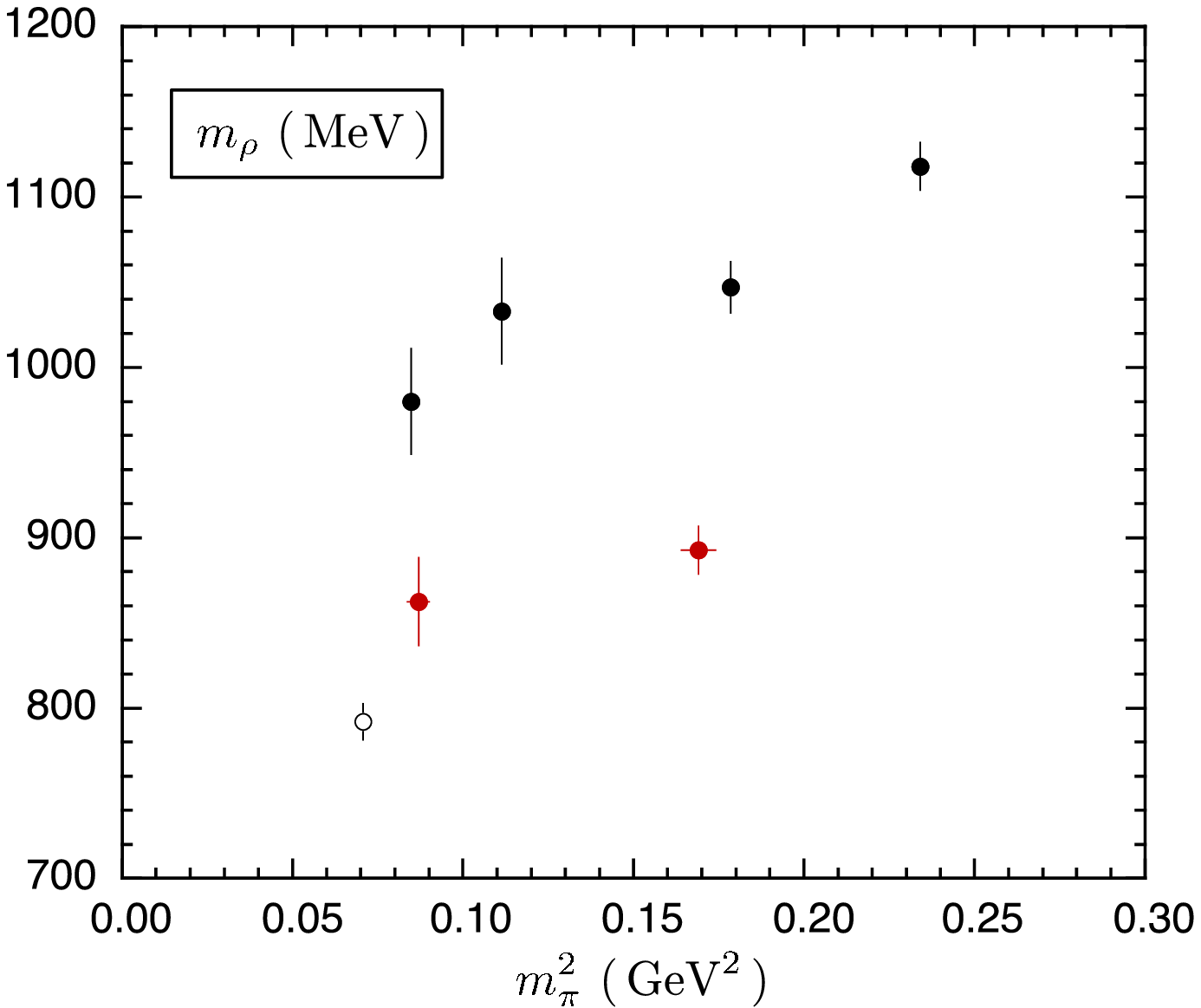}
\caption{
Comparison of our results (PACS-CS) obtained in $2+1$ flavor QCD
with those by ETMC
and Lang {\it et al.} in $2$ flavor QCD.
Left panel shows
the effective coupling constant $g_{\rho\pi\pi}$ and
right is the resonance mass $m_\rho$.
}
\label{fig:comp}
\end{figure}
%

In Fig.~\ref{fig:comp}
we compare our results (PACS-CS) obtained  in $2+1$ flavor QCD
with those by ETMC~\cite{rhd:SCPH:ETMC_1,rhd:SCPH:ETMC_2}
and Lang {\it et al.}~\cite{rhd:SCPH:LANG} in $2$ flavor QCD.
The left panel shows
the effective coupling constant $g_{\rho\pi\pi}$
and the right panel displays the resonance mass $m_\rho$
as a function of $m_\pi^2$.
A good agreement between our result and ETMC is
observed for $g_{\rho\pi\pi}$.
The result for the coupling constant
by Lang {\it et al.} takes a slightly smaller value,
but it is almost consistent with other works.

We see, however, large discrepancy for the resonance mass $m_\rho$
in the right panel of Fig.~\ref{fig:comp}.
The three groups worked at a single lattice spacing,
therefor a possible reason of
the discrepancy is the discretization error
due to the finite lattice spacing.
We can also consider several other reasons,
the reliability of the determination of the lattice spacing,
the dynamical strange quark effect,
the isospin breaking effect and so on,
but a definite conclusion can not be given here.
%
%
\section{ Summary }
We have reported on a calculation of the $P$-wave scattering phase shift
for the isospin $I=1$ two-pion system
and an estimation of the resonance parameters
from the $N_f=2+1$ full QCD configurations
with a large lattice volume.
The calculations are carried out at two quark masses,
which correspond to $m_\pi=410\,{\rm MeV}$ and $300\,{\rm MeV}$.
Our results of the effective coupling constant $g_{\rho\pi\pi}$
at the two quark masses are consistent within the
statistical error and also with the experiment.
This suggests a weak quark mass dependence
of the coupling constant.
We find a discrepancy for the resonance mass $m_\rho$
among three lattice studies.
Calculations near or on the physical point closer to the continuum limit
are necessary for a precise determination
of the resonance mass from lattice QCD.
We leave this issue to studies in the future.
%
%
\section*{Acknowledgments}
This work is supported in part by Grants-in-Aid
of the Ministry of Education
(Nos.
%
20340047, 20105001, 20105003,
%
20540248, 23340054,
%
21340049,
%
22244018, 20105002,
%
22105501, 22740138,
%
23540310,
%
22540265, 23105701,
%
10143538,
%
21105501, 23105708,
%
20105005 ).
The numerical calculations have been carried out
on PACS-CS at Center for Computational Sciences, University of Tsukuba.
%
%

%
%
\end{document}